\documentclass[epj]{svjour}
\usepackage{graphics}
\usepackage{amsmath,amsfonts,amssymb,epsfig,color}
\begin{document}
\title{Axial asymmetry in the IVBM}
\author{H. G. Ganev\inst{1} 
\thanks{On
leave of absence from the Institute of Nuclear Research and Nuclear
Energy, Bulgarian Academy of Sciences, Sofia, Bulgaria}%
}                     
\institute{Joint Institute for Nuclear Research, 141980 Dubna,
Moscow Region, Russia}
\date{Received: date / Revised version: date}
%
\abstract{The dynamical symmetry limit of the two-fluid Interacting
Vector Boson Model (IVBM), defined through the chain $Sp(12,R)
\supset U(3,3) \supset U_{p}(3) \otimes \overline{U_{n}(3)} \supset
SU^{\ast}(3) \supset SO(3)$, is considered and applied for the
description of nuclear collective spectra exhibiting axially
asymmetric features. The effect of the introduction of a Majorana
interaction to the $SU^{\ast}(3)$ model Hamiltonian on the
$\gamma$-band energies is studied. The theoretical predictions are
compared with the experimental data for $^{192}Os$, $^{190}Os$, and
$^{112}Ru$ isotopes. It is shown that by taking into account the
full symplectic structures in the considered dynamical symmetry of
the IVBM, the proper description of the energy spectra and the
$\gamma$-band energy staggering of the nuclei under considerations
can be achieved. The obtained results show that the potential energy
surfaces for the following two nuclei $^{192}Os$ and $^{112}Ru$,
possess almost $\gamma$-flat potentials with very shallow triaxial
minima, suggesting a more complex and intermediate situation between
$\gamma$-rigid and $\gamma$-unstable structures. Additionally, the
absolute $B(E2)$ intraband transition probabilities between the
states of the ground state band and $\gamma$ band, as well as the
$B(M1)$ interband transition probabilities between the states of the
ground and $\gamma$ bands for the two nuclei $^{192}Os$ and
$^{190}Os$ are calculated and compared with experiment and for the
$B(E2)$ values with the predictions of some other collective models
incorporating the $\gamma$-rigid or $\gamma$-unstable structures.
The obtained results agree well with the experimental data and
reveal the relevance of the used dynamical symmetry of IVBM in the
description of nuclei exhibiting axially asymmetric features in
their spectra.
\PACS{
      {21.60.Fw}{Models based on group theory}   \and
      {21.10.Re}{Collective levels}
     } 
} 
\maketitle

\section{Introduction}

It has been known for a long time that in certain mass regions
nuclei with static deformation show deviations from a rigid axially
symmetric picture. The possibility of static triaxial shapes for the
ground state of nuclei is a long-standing problem in nuclear
structure physics despite the fact that very few candidates have
been found experimentally \cite{ZC1},\cite{SJ}. In the geometrical
approach the triaxial nuclear properties are usually interpreted in
terms of either the $\gamma$-unstable rotor model of Wilets and Jean
\cite{WJ} or the rigid triaxial rotor model (RTRM) of Davydov
\emph{et al.} \cite{DF}. These models exploit the geometrical
picture of nucleus according to the Collective Model of Bohr and
Mottelson, expressed  in terms of the intrinsic variables $\beta$
and $\gamma$ where the former specifies the ellipsoidal quadrupole
deformation and the latter the degree of axial asymmetry. To
describe the deviations from axial symmetry the model of Wilets and
Jean assumes that the potential energy is independent of the
$\gamma$-degree of freedom, while in the model of Davydov \emph{et
al.} one considers a harmonic oscillator potential with a minimum at
finite values of $\gamma$ producing a rigid triaxial shape of the
nucleus.

The question of whether asymmetric atomic nuclei are
$\gamma$-unstable or $\gamma$-rigid has been an ongoing and active
issue in nuclear structure physics for over half a century.  A
number of signatures of $\gamma$-unstable and $\gamma-$rigid
structures in nuclei has been discussed
\cite{ZC1},\cite{SJ},\cite{SSa}-\cite{SSd}. While it might be
thought that the potential energy surfaces that are nearly
$\gamma$-flat or which display deep minima for some value of
$\gamma$ would produce rather different nuclear spectra, this is in
fact not the case. Indeed, the predictions for $\gamma$-unstable and
$\gamma$-rigid potentials are nearly identical for most observables
if the average value of $\gamma$ in the first case, $\gamma_{rms}$,
is equal to the $\gamma_{rigid}$ in the second, a situation
occurring for example in the Os-Pt region. However, a clear
distinction arises in the $\gamma$ band, where both the
$\gamma$-unstable and $\gamma$-rigid models exhibit an opposite
energy staggering. The comparison of a $\gamma$-rigid rotor and a
$\gamma$-unstable models yields similar ground state band energies,
but the levels of $\gamma$-band are grouped as $2^{+}$, $(3^{+},
4^{+})$, $(5^{+}, 6^{+})$, $\ldots$ in $\gamma$-unstable and as
$(2^{+}, 3^{+})$, $(4^{+}, 5^{+})$, $\ldots$ in $\gamma$-rigid
models, respectively. Thus, obviously the structure of the $\gamma$
band is crucial for the identification of the shape in the real
nuclei and hence for the manifestation of the $\gamma$ degrees of
freedom.

In Ref.\cite{TSIVBM} a new dynamical symmetry limit of the two-fluid
Interacting Vector Boson Model (IVBM) was introduced, which seems to
be appropriate for the description of deformed even-even nuclei,
exhibiting triaxial features. It was shown there that the addition
of Majorana interaction to the $SU^{\ast}(3)$ model Hamiltonian
produces a stable triaxial minimum. In this paper, we develop
further our theoretical approach initiated in Ref.\cite{TSIVBM} by
considering in more details the spectra of some even-even
transitional nuclei, supposed to be axially asymmetric, in the
framework of the symplectic IVBM with $Sp(12,R)$ as a group of
dynamical symmetry. We focus on the $\gamma$-band properties and
show how the $\gamma$-band energies (and the corresponding energy
staggering) are affected by the presence of the introduced
interaction. The theoretical predictions are compared with the
experimental data for $^{192}Os$, $^{190}Os$, and $^{112}Ru$
isotopes. It will be shown that by taking into account the full
symplectic structures in the considered dynamical symmetry of the
IVBM, the proper description of the energy spectra and the
$\gamma$-band energy staggering of the nuclei under considerations
can be achieved.

Additionally, the absolute $B(E2)$ intraband transition
probabilities between the states of the ground state band and
$\gamma$ band, as well as the $B(M1)$ interband transition
probabilities between the states of the ground and $\gamma$ bands
for the two nuclei $^{192}Os$ and $^{190}Os$ are considered and
compared with experiment and for the $B(E2)$ values with the
predictions of some other collective models incorporating the
$\gamma$-rigid or $\gamma$-unstable structures.

\section{The algebraic structure of the $U(3,3)$ dynamical symmetry}

It was suggested by Bargmann and Moshinsky
\cite{BargMosh1},\cite{BargMosh2} that two types of bosons are
needed for the description of nuclear dynamics. It was shown there
that the consideration of only two-body system consisting of two
different interacting vector particles will suffice to give a
complete description of $N$ three-dimensional oscillators with a
quadrupole-quadrupole interaction. The latter can be considered as
the underlying basis in the algebraic construction of the
\emph{phenomenological} IVBM.

The algebraic structure of the IVBM \cite{PSIVBM} is realized in
terms of creation and annihilation operators of two kinds of vector
bosons $u_{m}^{\dag}(\alpha )$, $u_{m}(\alpha )$ ($m=0,\pm 1$),
which differ in an additional quantum number $\alpha=\pm1/2$ (or
$\alpha=p$ and $n$)$-$the projection of the $T-$spin (an analogue to
the $F-$spin of IBM-2 or the $I-$spin of the particle-hole IBM). In
the present paper, we consider these two bosons just as elementary
building blocks or quanta of elementary excitations (phonons) rather
than real fermion pairs, which generate a given type of algebraic
structures. Thus, only their tensorial structure is of importance
and they are used as an auxiliary tool, generating an appropriate
\emph{dynamical} symmetry.

We consider the following reduction chain of the dynamical symmetry
group $Sp(12,R)$ of the IVBM for studying the triaxiality in atomic
nuclei:
\begin{equation}
\begin{tabular}{lllll}
$Sp(12,R)$ & $\supset $ & $U(3,3)$ &  &  \\
&  &  $\ \ \ \ \nu $ &  &  \\
&  &  &  &  \\
& $\supset $ & $U_{p}(3)$ & $\otimes $ & $\overline{U_{n}(3)}$ \\
&  & $[N_{p}]_{3}$ &  & $[-N_{n}]_{3}$ \\
&  &  &  &  \\
& $\supset $ & $SU^{\ast }(3)$ & $\supset $ & $SO(3),$ \\
&  &  $(\lambda ,\mu )$ & $K$ & $\ \ L$%
\end{tabular}
\label{NDS}
\end{equation}
where the labels below the different subgroups are the quantum
numbers corresponding to their irreducible representations (irreps).
As it was shown in Ref. \cite{TSIVBM}, this dynamical symmetry is
appropriate for nuclei in which the one type of particles is
particle-like and the other is hole-like. For more details
concerning the algebraic structure of this dynamical symmetry of the
IVBM, we refer the readers to Ref.\cite{TSIVBM}.

All bilinear combinations of the creation and annihilation operators
of the two vector bosons generate the boson representations of the
non-compact symplectic group $ Sp(12,R)$:
\begin{eqnarray}
F_{M}^{L}(\alpha ,\beta ) = {\sum }_{k,m}C_{1k1m}^{LM}u_{k}^{+}(
\alpha )u_{m}^{+}(\beta ), \label{Fs} \\
G_{M}^{L}(\alpha ,\beta ) ={\sum }_{k,m}C_{1k1m}^{LM}u_{k}(\alpha
)u_{m}(\beta ), \label{Gs} \\
A_{M}^{L}(\alpha, \beta )={\sum }_{k,m}C_{1k1m}^{LM}u_{k}^{+}(\alpha
)u_{m}(\beta ),  \label{numgen}
\end{eqnarray}
where $C_{1k1m}^{LM}$, which are the usual Clebsch-Gordan
coefficients for $L=0,1,2$ and $M=-L,-L+1,...L$, define the
transformation properties of (\ref{Fs}),(\ref{Gs}) and
(\ref{numgen}) under rotations. We also introduce the following
notations $u_{m}^{\dag}(\alpha=1/2 )=p^{\dag}_{m}$ and
$u_{m}^{\dag}(\alpha=-1/2 )=n^{\dag}_{m}$. In terms of the $p-$ and
$n-$boson operators, the Weyl generators of the ladder
representation of $U(3,3)$ are \cite{TSIVBM}
\begin{equation}
p_{k}^{\dag}p_{m}, \ \ \ \ p_{k}^{\dag}n_{m}^{\dag}, \ \ \ \ \
-n_{k}p_{m}, \ \ \ \ -n_{m}^{\dag}n_{k},
 \label{BRU33}
\end{equation}
which are obviously a subset of symplectic generators
(\ref{Fs})$-$(\ref{numgen}). The first-order Casimir operator of
$U(3,3)$ is
\begin{equation}
C_{1}[U(3,3)]=\sum_{k}(p_{k}^{\dag}p_{k}-n_{k}^{\dag}n_{k}),
 \label{FCOU33}
\end{equation}
and does not differ essentially from the operator $T_{0}$ defined in
\cite{PSIVBM}:
\begin{equation}
T_{0}=\frac{1}{2}C_{1}[U(3,3)]+\frac{3}{2}.
\end{equation}
The $U(3,3)$ irreps (ladder irreducible representations) contained
in either (even) $<(1/2)^{6}>$ or (odd) $<(1/2)^{5}3/2>$ irrep of
$Sp(12,R)$ can be denoted by the shorthand notation $[\nu]$, $\nu
\in Z$, defined as follows \cite{TSIVBM}:
\begin{align}
[\nu]&=\{(1/2)^{2},\nu+\frac{1}{2};(1/2)^{3}\} \ \ \ \ \emph{if}\ \ \nu > 0 \label{nu1} \\
&=\{(1/2)^{3};(1/2)^{2},-\nu+\frac{1}{2}\} \ \ \emph{if}\ \ \nu < 0 \label{nu2} \\
&=\{(1/2)^{3};(1/2)^{3}\}, \ \ \ \ \ \ \ \ \ \ \ \  \emph{if}\ \ \nu
= 0 \label{nu3}
\end{align}
The branching rules can be written as
\begin{equation}
<(1/2)^{6}> \ \ \downarrow \sum_{{\nu=-\infty},{\nu=even}}^{+\infty}
\oplus[\nu] \label{EIR}
\end{equation}
and
\begin{equation}
<(1/2)^{5}3/2> \ \ \downarrow
\sum_{{\nu=-\infty},{\nu=odd}}^{+\infty} \oplus[\nu]. \label{OIR}
\end{equation}
In the present application we consider only the even irreducible
representation of $Sp(12,R)$.

The direct product $U_{p}(3) \otimes \overline{U_{n}(3)}$ subalgebra
is defined by the subset of the number preserving generators
($\ref{BRU33}$) of $U(3,3)$, namely
\begin{equation}
p_{k}^{\dag}p_{m},  \ \ \ \ -n_{m}^{\dag}n_{k}.
 \label{U3U3}
\end{equation}
Then, the combined (particle-hole) algebra $U^{\ast}(3)$ is simply
expressed by the linear combination operators $A_{km} \equiv
\{p_{k}^{\dag}p_{m}-n_{m}^{\dag}n_{k}\}$ of ($\ref{U3U3}$), which
can also be defined in the following way \cite{TSIVBM}
\begin{align}
&M=N_{p}-N_{n}, \label{M} \\
&L_M = L^{p}_{M}+L^{n}_{M}, \label{LM} \\
&Q_M = Q^{p}_{M}-Q^{n}_{M}. \label{Qminus}
\end{align}
The second order Casimir operator of $U^{\ast}(3)$ can be defined by
\begin{equation}
C_{2}[U^{\ast}(3)]=\sum_{ij}A_{ij}A_{ji}. \label{C2U*3}
\end{equation}
The $SU^{\ast}(3)$ algebra is obtained by excluding the operator
($\ref{M}$) which is the single generator of the $O(2)$ algebra,
whereas the angular momentum algebra $SO(3)$ is generated by the
generators $L_{M}$ only.

We can label the basis states according to the chain (\ref{NDS}) as:
\begin{eqnarray}
|\nu;N_{p},N_{n};(\lambda,\mu);KL ~\rangle,  \label{Basis}
\end{eqnarray}
where $\nu$ is the eigenvalue of the $U(3,3)$ first order Casimir
operator,  $N_{p}$ and $N_{n}$ label the $U_{p}(3) \otimes
\overline{U_{n}(3)}$ irreps  ,$(\lambda,\mu)$ are the $SU^{\ast}(3)$
quantum numbers, $K$ is the multiplicity index in the reduction
$SU(3)\supset SO(3)$, and $L$ is the angular momentum of the
corresponding collective state.

The basis states associated with the even irreducible representation
of the $Sp(12,R)$ can be constructed by the application of powers of
raising generators $F_{M}^{L}(\alpha ,\beta )$ of the same group on
the boson vacuum state. Each raising operator will increase the
number of bosons $N$ by two. The $Sp(12,R)$ classification scheme
for the $SU^{\ast}(3)$ boson representations is shown on Table
\ref{BS}. The ladder representations of the non-compact algebra
$U(3,3)$ act along the columns ("ladders") in the space of the boson
representation of the $Sp(12,R)$ algebra, defined through the
eigenvalues  $[\nu]$ of the first Casimir operator ($\ref{FCOU33}$)
of the $U(3,3)$ algebra. There exists a connection between this
ladder representation ("vertical classification") and the boson
representation of $U(6) \subset Sp(12,R)$ ("horizontal
classification"). Each row (fixed $N$) of the table corresponds to a
given irreducible representation of the $U(6)$. Note that the number
of bosons $N=N_{p}+N_{n}$ is not a good quantum number along the
ladder representations of $U(3,3)$.

\begin{table}[h]
\caption{Symplectic classification of the $SU^{\ast}(3)$ basis
states.} \label{BS}
\smallskip\centering\small\addtolength{\tabcolsep}{-4.5pt}
\begin{tabular}{||l||llll|l|llll||}
\hline\hline $N\backslash \nu $ & $\cdots $ &
\multicolumn{1}{|l}{$6$} & \multicolumn{1}{|l}{$4$} &
\multicolumn{1}{|l|}{$2$} & \ $\ \ 0$ & $-2$ &
\multicolumn{1}{|l}{$-4$} & \multicolumn{1}{|l}{$-6$} &
\multicolumn{1}{|l||}{$\cdots $} \\ \hline\hline $0$ &  &  &  &  &
$\ (0,0)$ &  &  &  &  \\ \cline{1-1}\cline{5-7}
$2$ &  &  &  & \multicolumn{1}{|l|}{%
\begin{tabular}{l}
$(2,0)$ \\
\ \
\end{tabular}%
} &
\begin{tabular}{l}
$(1,1)$ \\
$(0,0)$%
\end{tabular}
&
\begin{tabular}{l}
$(0,2)$ \\
\ \ \
\end{tabular}
& \multicolumn{1}{|l}{} &  &  \\ \cline{1-1}\cline{4-8}
$4$ &  &  & \multicolumn{1}{|l}{%
\begin{tabular}{l}
$(4,0)$ \\
\\
\ \ \
\end{tabular}%
} & \multicolumn{1}{|l|}{%
\begin{tabular}{l}
$(3,1)$ \\
$(2,0)$ \\
\
\end{tabular}%
} &
\begin{tabular}{l}
$(2,2)$ \\
$(1,1)$ \\
$(0,0)$%
\end{tabular}
&
\begin{tabular}{l}
$(1,3)$ \\
$(0,2)$ \\
\
\end{tabular}
& \multicolumn{1}{|l}{%
\begin{tabular}{l}
$(0,4)$ \\
\\
\
\end{tabular}%
} & \multicolumn{1}{|l}{} &  \\ \cline{1-1}\cline{3-9}
$6$ &  & \multicolumn{1}{|l}{%
\begin{tabular}{l}
$(6,0)$ \\
\\
\\
\
\end{tabular}%
} & \multicolumn{1}{|l}{%
\begin{tabular}{l}
$(5,1)$ \\
$(4,0)$ \\
\\
\
\end{tabular}%
} & \multicolumn{1}{|l|}{%
\begin{tabular}{l}
$(4,2)$ \\
$(3,1)$ \\
$(2,0)$ \\
\
\end{tabular}%
} &
\begin{tabular}{l}
$(3,3)$ \\
$(2,2)$ \\
$(1,1)$ \\
$(0,0)$%
\end{tabular}
&
\begin{tabular}{l}
$(2,4)$ \\
$(1,3)$ \\
$(0,2)$ \\
\
\end{tabular}
& \multicolumn{1}{|l}{%
\begin{tabular}{l}
$(1,5)$ \\
$(0,4)$ \\
\\
\
\end{tabular}%
} & \multicolumn{1}{|l}{%
\begin{tabular}{l}
$(0,6)$ \\
\\
\\
\
\end{tabular}%
} & \multicolumn{1}{|l||}{} \\ \cline{1-1}\cline{3-9} $\vdots $ &  &
\multicolumn{1}{|l}{$\ \ \ \ \ \vdots $} & \multicolumn{1}{|l}{$\ \
\ \ \vdots $} & \multicolumn{1}{|l|}{$\ \ \ \ \vdots $} & $\ \ \ \
\vdots $ & $\ \ \ \ \ \vdots $ & \multicolumn{1}{|l}{$\ \ \ \ \vdots
$} & \multicolumn{1}{|l}{$\ \ \ \ \vdots $} &
\multicolumn{1}{|l||}{}%
\end{tabular}
\end{table}

\section{The energy spectrum}

The most general Hamiltonian with $SU^{\ast}(3)$ symmetry consists
of the Casimir invariants of $SU^{\ast}(3)$ and its subgroup $SO(3)$
\begin{equation}
H=aC_{2}[SU^{\ast}(3)]+bC_{2}[SO(3)], \label{HSU*3}
\end{equation}
where
\begin{equation}
C_{2}[SU^{\ast}(3)]=\tfrac{1}{6}Q^2 +\tfrac{1}{2}L^{2}
\end{equation}
and the quadrupole operator Q is given by (\ref{Qminus}).

The spectrum of this Hamiltonian is determined by
\begin{equation}
E=a(\lambda^{2}+\mu^{2}+\lambda\mu+3\lambda+3\mu)+bL(L+1).
\label{ESU*3}
\end{equation}
We point out that there are very large degeneracies in the resulting
energy spectrum caused by the large values of $\lambda$ and $\mu$,
which is not observed in the real nuclear spectra.

In the present application we consider $Sp(12,R)$ to be the group of
the dynamical symmetry of the model. We make use of the following
Hamiltonian:
\begin{align}
H_{U(3,3)}=&a_{1}M^{2} + b(N_{n}^{2}-N_{p}^{2}) \notag \\
&+ a_{3}C_{2}[SU^{\ast}(3)] + b_{3}C_{2}[SO(3)], \label{HU33}
\end{align}
expressed as a linear combination of the Casimir operators of the
different subgroups in the chain (\ref{NDS}). The Hamiltonian
(\ref{HU33}) is diagonal in the basis (\ref{Basis}). Then its
eigenvalues that yield the spectrum of the nuclear systems are:
\begin{align}
&E(\nu;N_{p},N_{n};(\lambda,\mu);L)=a_{1}\nu^{2} + b(N_{n}^{2}-N_{p}^{2}) \notag \\
&+ a_{3}(\lambda^{2}+\mu^{2}+\lambda\mu+3\lambda+3\mu) +
b_{3}L(L+1). \label{EU33}
\end{align}
The energy spectrum determined by Eq.(\ref{EU33}) will be the
starting point for our further calculations.

\section{Triaxial shapes in the IVBM}

In Ref.\cite{TSIVBM} it has been shown that the addition of
different types of perturbations to the $SU^{\ast}(3)$ energy
surface, in particular the addition of a Majorana interaction and an
$O(6)$ term to the $SU^{\ast}(3)$ model Hamiltonian, produces a
stable triaxial minimum in the potential energy surfaces. In present
work we consider only the  inclusion of a Majorana interaction to
the model Hamiltonian and study the influence of the latter on the
produced low-lying energy spectra. We expect that many experimental
properties of some deformed even-even nuclei, exhibiting axially
asymmetric features, to be explained with the perturbed Hamiltonian
under consideration.

We present a schematic calculations starting with the Hamiltonian H
(\ref{HU33}) to which a perturbation term is added. The Hamiltonian
which contains Majorana interaction is written in the form
\begin{equation}
H= H_{U(3,3)} + aM_{3}, \label{HpM3}
\end{equation}
where the Majorana operator is defined as
\begin{equation}
M_{3}=2(p^{\dag} \times n^{\dag})^{(1)} \cdot (p \times n)^{(1)}
\label{M3}
\end{equation}
and it is related to the $U(3)$ second order Casimir invariant
$C_{2}[U(3)]$, defined in Ref.\cite{PSIVBM}, via the relation
\begin{equation}
C_{2}[U(3)]=N(N+2)-2M_{3}. \label{M3C2}
\end{equation}
The Hamiltonian H contains the pure $SU^{\ast}(3)$ symmetry, when
only $a_{3} \neq 0$ in Eq.(\ref{HU33}). The Majorana interaction can
be written in the form
\begin{align}
M_{3}&=\frac{2}{3}(N^{2}_{p}+N^{2}_{n})+N_{p}+N_{n}-\frac{1}{3}M^{2}-\frac{1}{2}C_{2}[SU^{\ast}(3)] \notag \\
&-\frac{1}{3} \Bigg(\frac{15}{64}(C_{2}[SU^{\ast}(3)]-C_{2}[SU_{p}(3)]-C_{2}[\overline{SU_{n}(3)}]) \notag \\
&-\frac{3}{32}\sqrt{30}[A^{(1,1)}(p,p) \times
A^{(1,1)}(n,n)]^{(2,2)}\Bigg).  \label{M3ND}
\end{align}
It is evident that the only non-diagonal component of the Majorana
interaction along the chain (\ref{NDS}) is represented by the last
term. It mixes different $SU(3)$ irreps.

In our application, the most important point is the identification
of the experimentally observed states with a certain subset of basis
states from symplectic extension of the model (Table \ref{BS}). As
in our previous applications of the symplectic IVBM, we use the
algebraic concept of \textquotedblleft yrast\textquotedblright\
states, introduced in \cite{GGG}. According to this concept we
consider as yrast states the states with given $L$ that minimize the
energy with respect to the number of vector bosons $N$ that build
them. Since, the GSB in the triaxial nuclei is supposed to belong to
the $SU^{\ast}(3)$ irreps of the type $(\lambda=N_{p},\mu=N_{n})$,
we map the states of the GSB onto the ladder representation of
$U(3,3)$ with $\nu=0$ (middle column of Table \ref{BS}). The
presented mapping of the experimental states onto the $SU(3)$ basis
states, using the algebraic notion of yrast states, is a particular
case of the so called "stretched" states \cite{str}, which in our
case are defined as the states of the type $(\lambda,\mu
)=(\lambda_{0}+k,\mu_{0}+k)$, where $k=0,2,4, \ldots$. In the
symplectic extension of the IVBM the change of the number $k$, which
is related in the applications to the angular momentum $L$ of the
states, gives rise to the collective bands. Thus, explicitely the
states of the GSB are identified with the $SU^{\ast}(3)$ multiplets
$(\lambda,\mu )=(k,k)$, where $k=L$. The same type of stretched
states $(\lambda_{0}+k,\mu_{0}+k)$ are associated with the states
from the $\gamma$ band, where the symplectic band head structure of
the considered band is determined by the initial number of phonons
$N_{0}=\lambda_{0} + \mu_{0}=6$ ($\lambda_{0}=2$, $\mu_{0}=4$).
Additionally, for the $\gamma$ band to each single $SU^{\ast}(3)$
irrep $(\lambda,\mu)$ ($k$-fixed) we put into correspondence two
consecutive states with angular momentum $L$ and $L+1$,
respectively. This choice allows us to reproduce the doublet
structure of the $\gamma$-band. We note that the present choice of
the  $SU^{\ast}(3)$ multiplets associated with the states of the
$\gamma$-band is quite similar to the phonon multiplet structure of
the $\gamma$-band states within the framework of the IBM-1 in its
$O(6)$ limit, where the states cluster in doublets differing in the
$O(5)$ label $\tau$, which corresponds to the phonon-like quantum
number $\Lambda$ in the $\gamma$-unstable model of Wilets and Jean
\cite{ZC1}.

\begin{figure}[h]\centering
\includegraphics[width=80mm]{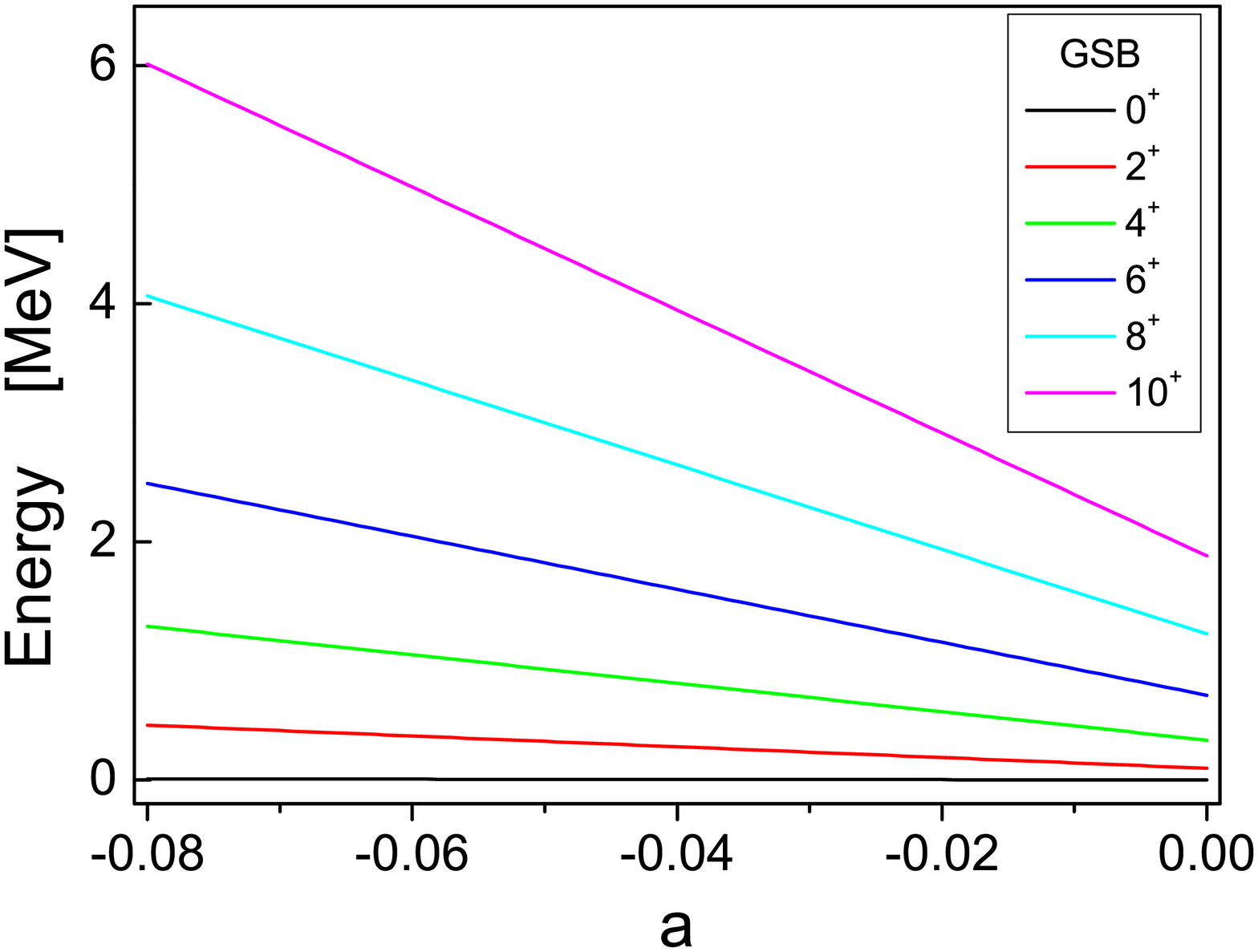}\hspace{1.mm}
\includegraphics[width=80mm]{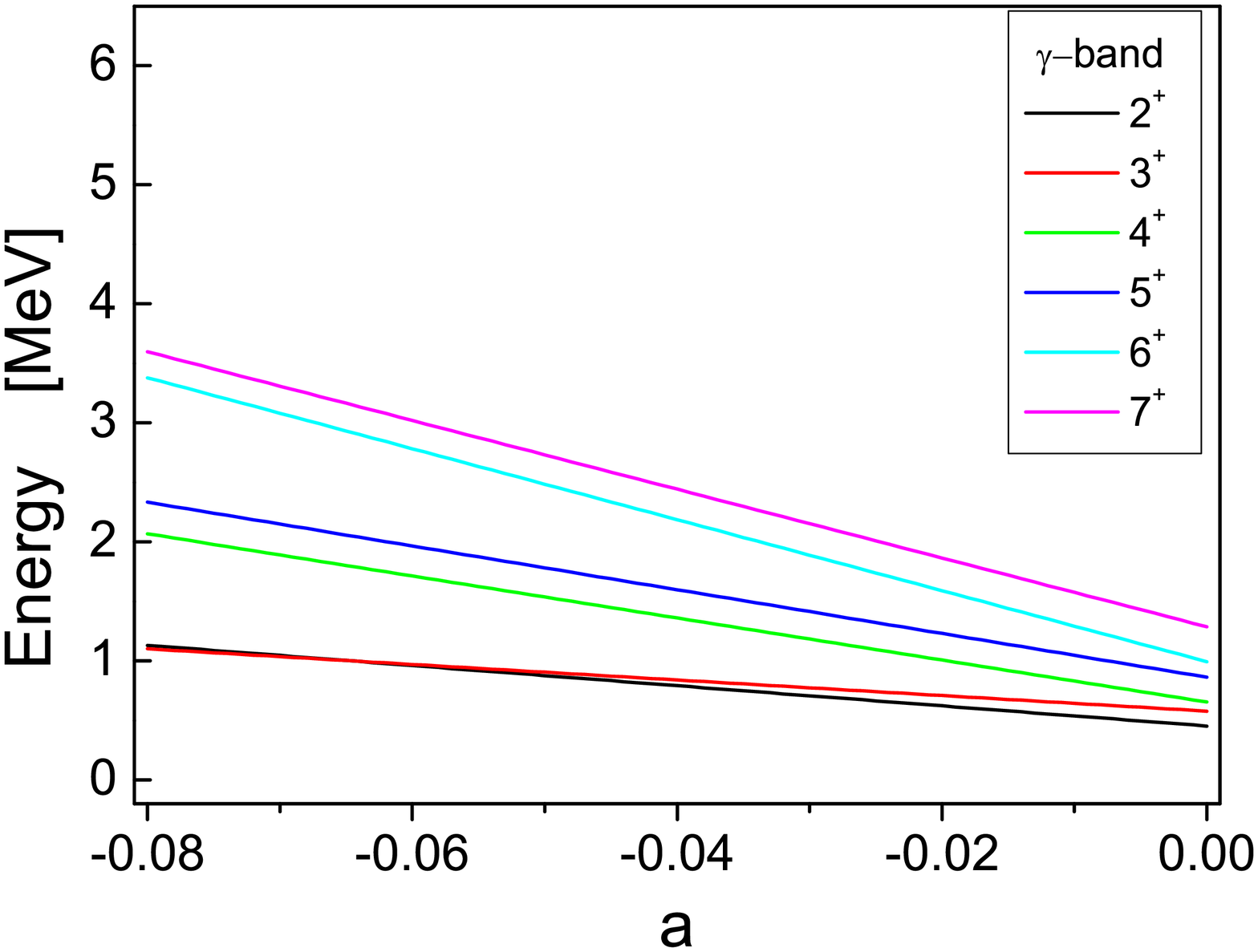}
\caption{(Color online) Energies of the ground and $\gamma$ bands as
a function of the strength parameter $a$. The values of the rest
model parameters are $a_{1}=0.10343$ MeV, $b = -0.00274$ MeV, $a_{3}
= -0.00116$ MeV and $b_{3}= 0.02092$ MeV.} \label{EHpM}
\end{figure}

To show the influence of the Majorana interaction on the energy
spectrum, we present the model calculations with the IVBM
Hamiltonian (\ref{HpM3}) in which the Majorana term is included and
diagonalyzed numerically. The evolution of the ground and $\gamma$
bands as a function of the strength parameter $a$ is shown in Fig.
1. From the figure one can see that the inclusion of the Majorana
term does not change the level spacings of the ground state band and
hence preserves its character. It can be also seen that the
$\gamma$-rigid-like doublet structure of the $\gamma$-band is
conserved for a wide interval of negative values of the parameter
$a$, but for $a = 0$ MeV (no mixing of the $SU(3)$ irreps) one
obtains the well known $\gamma$-unstable-like structure. For
$a\simeq -0.005$ MeV we obtain an intermediate situation with more
regular spacing of the energy levels.

\section{Numerical results}

\subsection{Energy spectra}

\begin{figure}[h]\centering
\includegraphics[width=80mm]{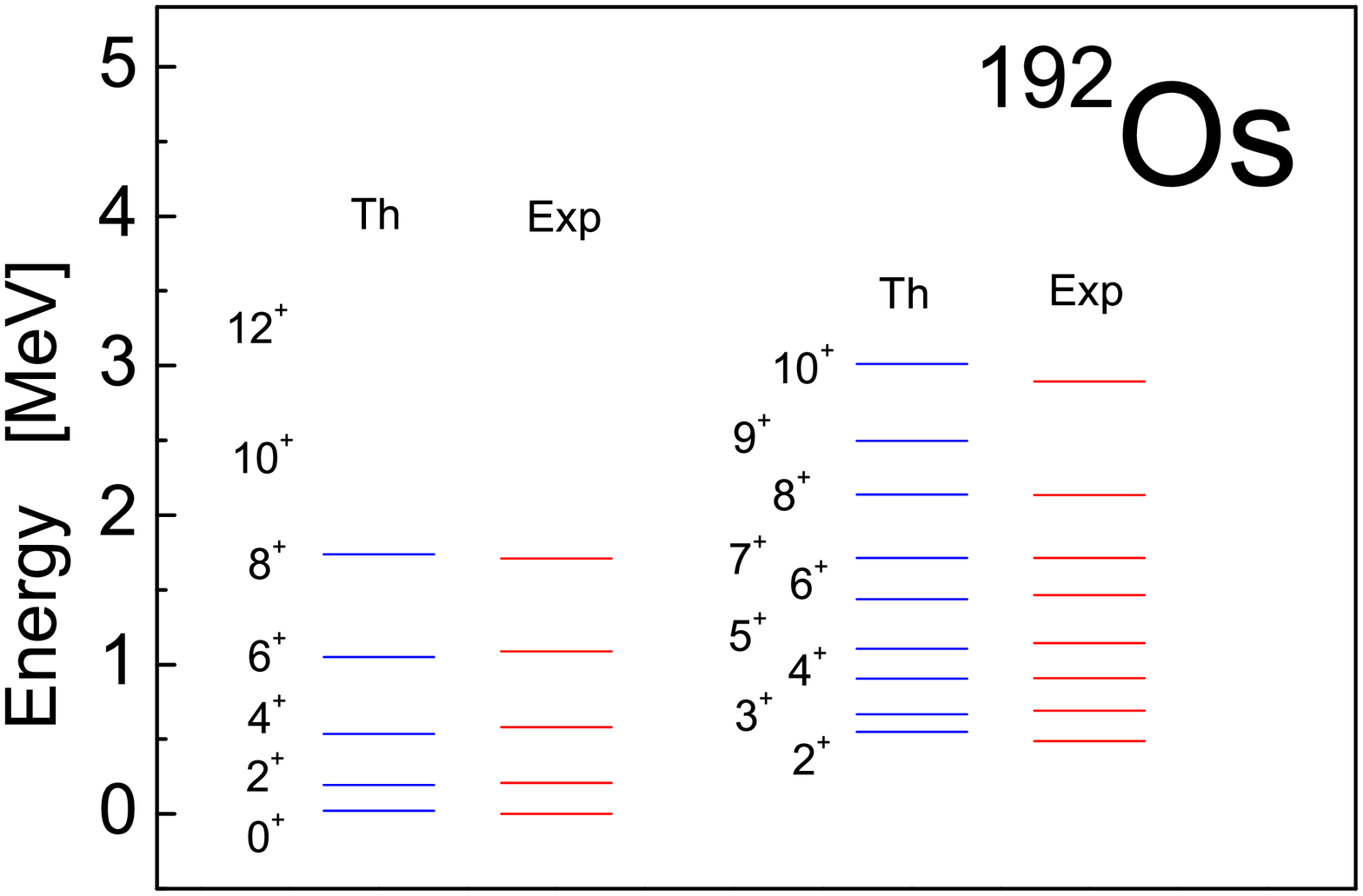}
\caption{(Color online) Excitation energies for GSB and $\gamma$
band of  $^{192}Os$. The values of the rest model parameters are
$a_{1} = 0.1034$ MeV, $b = -0.0027$ MeV, $a_{3} = -0.0011$ MeV,
$b_{3} = 0.0209$ MeV,  and $a = -0.0150$ MeV.} \label{Os192}
\end{figure}

\begin{figure}[h]\centering
\includegraphics[width=80mm]{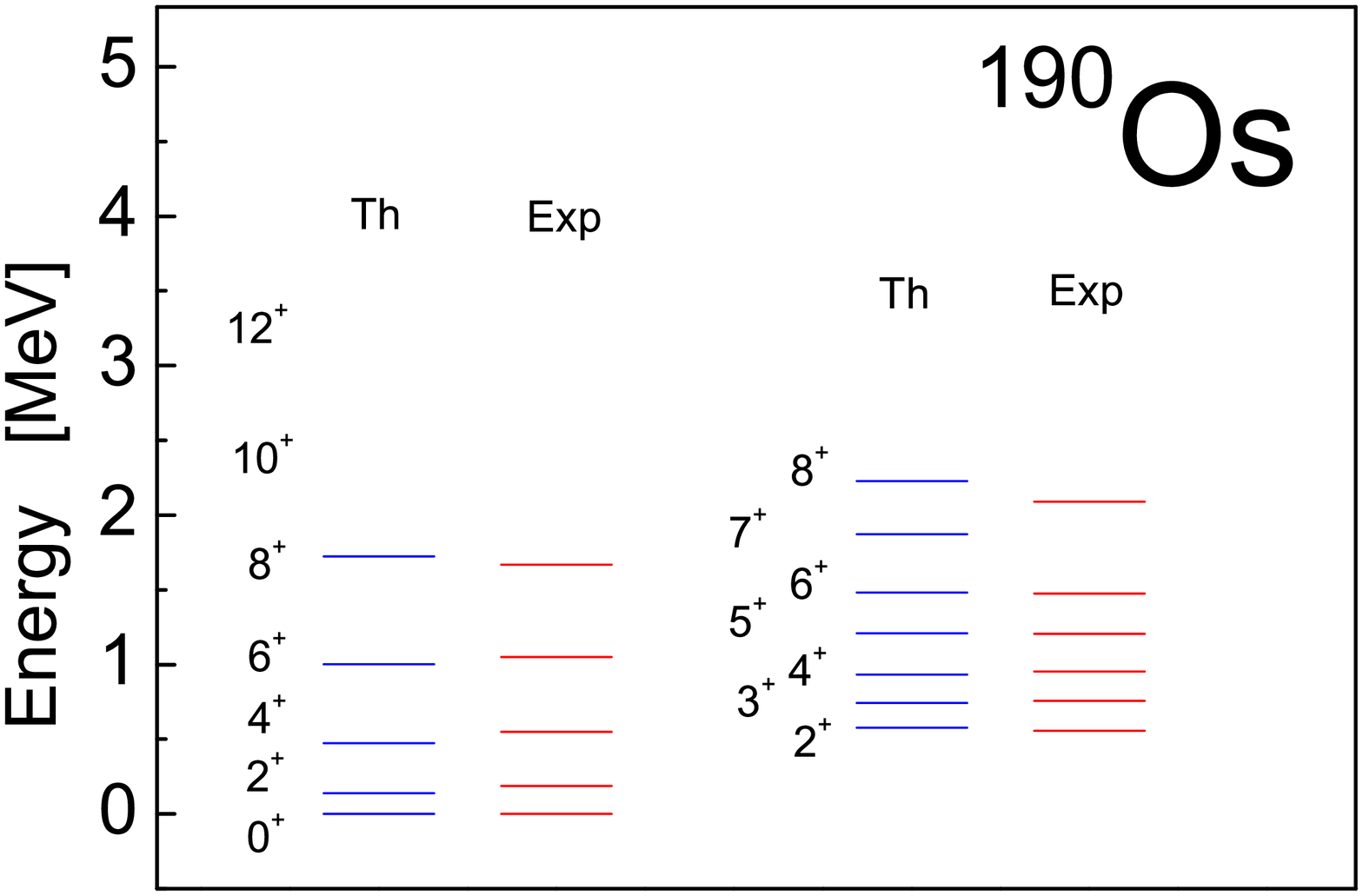}
\caption{(Color online) Excitation energies for GSB and $\gamma$
band of $^{190}Os$. The values of the rest model parameters are
$a_{1} = 0.1127$ MeV, $b = 0.0004$ MeV, $a_{3} = -0.0002$ MeV,
$b_{3} = 0.0251$ MeV,  and $a = 0$ MeV.} \label{Os190}
\end{figure}

\begin{figure}[h]\centering
\includegraphics[width=80mm]{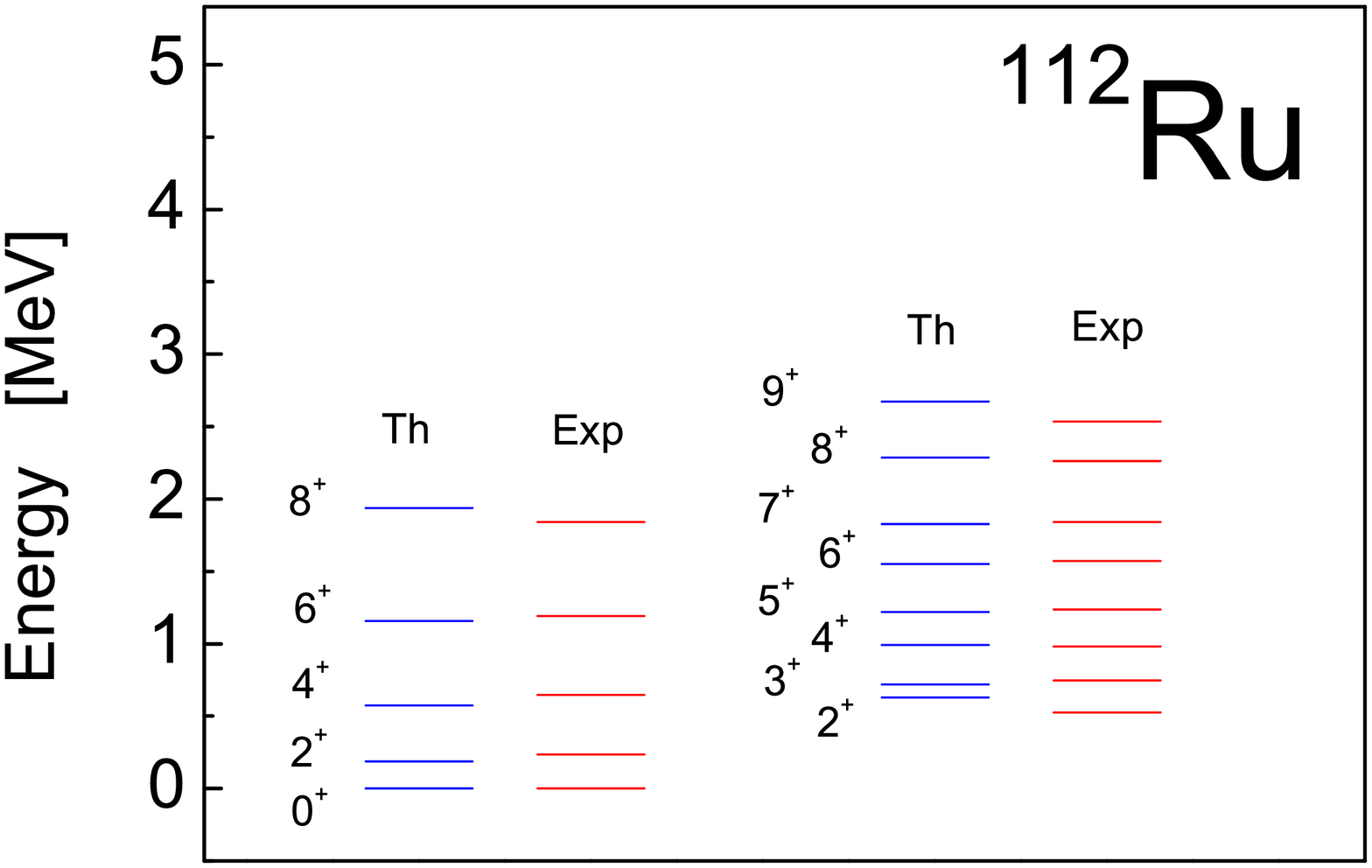}
\caption{(Color online) Excitation energies for GSB and $\gamma$
band of $^{112}Ru$. The values of the rest model parameters are
$a_{1} = 0.1144$ MeV, $b = -0.0056$ MeV, $a_{3} = -0.0014$ MeV,
$b_{3} = 0.0212$ MeV,  and $a = -0.0209$ MeV.} \label{Ru112}
\end{figure}

Our theoretical considerations are applied for the calculation of
the excitation energies of the ground and $\gamma$ bands in
$^{192}Os$, $^{190}Os$, and $^{112}Ru$, which are assumed in the
literature to possess triaxial shapes. We determine the values of
the model parameters $a_{1},b, \alpha_{3},\beta_{3}$, and $a$ by
fitting the energies of the ground and $\gamma$ bands for the
corresponding isotopes to the experimental data \cite{exp}, using a
$\chi^{2}$ procedure. The theoretical predictions, compared with
experiment, are presented in Figures \ref{Os192}$-$\ref{Ru112}.

There is a long-standing debate about the nature of the spectra of
$Os$ isotopes. Some groups consider these nuclei as being
$O(6)$-like with $\gamma$-unstable or $\gamma$-soft energy surfaces
with a prolate minimum \cite{NomuraOs1}, \cite{NomuraOs2}, while
other as asymmetric rotor \cite{DF},
\cite{OsTri1},\cite{OsTri2},\cite{OsTri3}, which assumes rigidity in
the $\gamma$ degrees of freedom. The Os isotopes considered here
have been treated in terms of the IBM in the transition region from
the rotor to the $\gamma$-unstable limit \cite{OsIBM}. In
Ref.\cite{Casten}, these isotopes are considered as a textbook
example of this transition. In Ref.\cite{VibGS1} it was shown that
the empirical deviations from the $O(6)$ limit of the IBM, in the
Os-Pt region, can be interpreted by introducing explicitly triaxial
degrees of freedom, suggesting a more complex and possibly
intermediate situation between $\gamma-$rigid and $\gamma-$unstable
properties. Indeed, as it can be seen from the presented examples,
the experimentally observed level spacings in the $\gamma$ band are
more regular. In terms of the potentials, this means that the true
potentials are $\gamma$-dependent.

Recently, a number of theoretical calculations \cite{Bonche},
\cite{CCQH}, \cite{Robledo}, \cite{Nomura2}, \cite{Nomura3} predict
a tiny region of triaxiality between the prolate and oblate shapes
in this mass region. The self-consistent Hartree-Fock-Bogoliubov
calculations \cite{Robledo} with Gogny D1S and Skyrme SLy4 forces
predict that the prolate to oblate transition takes place at neutron
number $N=116$ ($^{192}$Os).

The $^{96-108}Ru$ isotopes have also been described within the
framework of IBM-1 as transitional between $U(5)$ and $O(6)$ limits
\cite{IBMRu}, whereas in the Generalized Collective Model these
nuclei are described as transitional between spherical and triaxial
with a prolate onset for $^{96}Ru$ \cite{GCMRu}. The collective
structure of $^{104-112}Ru$ isotopes was also studied within the
framework of the RTRM \cite{RTRMRu1},\cite{RTRMRu2}. In the
$^{108,110,112}Ru$, the experimental E2 branching ratios were found
to be in overall agreement with those predicted by the RTRM
calculations, but the calculated odd-even staggering in the $\gamma$
band shows disagreement with experiment. It has been recognized that
in Ru nuclei the $\gamma$ dependence of the potential is
intermediate between the two limiting cases of a $\gamma$-unstable
and a $\gamma$-rigid rotor \cite{RTRMRu1}, \cite{RTRMRu2}.

From the Figs. \ref{Os192}$-$\ref{Ru112} one sees that the
calculated energy levels  of both ground state and $\gamma$ bands
agree rather well with the observed data and the doublet structure
of the $\gamma$ band, although slightly pronounced in experiment, is
correctly reproduced.

\subsection{The energy staggering}

A number of signatures of $\gamma$-unstable and $\gamma$-rigid
structures in nuclei has been discussed
\cite{ZC1},\cite{SJ},\cite{SSa}-\cite{SSd}. Many authors
investigated the transition from the $\gamma$-unstable regime to a
triaxial behavior. The two nuclear phases, as was mentioned, are
characterized by different doublet structures in the $\gamma$ band.
A useful quantity that distinguishes these two cases is the energy
staggering signature \cite{ZC1},\cite{SJ}:
\begin{equation}
S(L)=\frac{[E(L)-E(L-1)]-[E(L-1)-E(L-2)]}{E(2^{+}_{g})},
\label{Stag}
\end{equation}
where $E(L)$ stands for the energy of the state $L^{+}$ belonging to
the $\gamma$ band. The quantity $S(L)$ measures the displacement of
the $(L-1)^{+}$ level relative to the average of its neighbors,
$L^{+}$ and $(L-2)^{+}$, normalized to the energy of the first
excited state of the ground band, $2^{+}_{g}$. The doublet structure
is reflected in the sawtooth shape of the function $S(L)$.

\begin{figure}[h]\centering
\includegraphics[width=80mm]{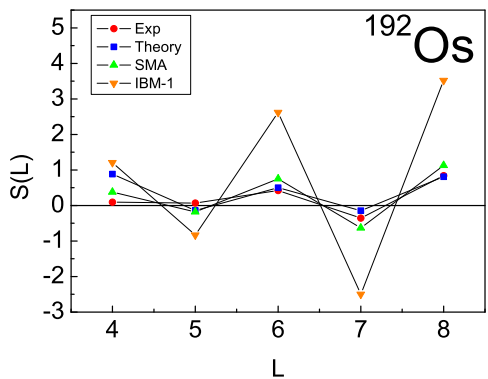}
\caption{(Color online) Calculated and experimental staggering of
the $\gamma$ band in $^{192}Os$. The predictions of the sextic and
Mathieu approach (SMA) \cite{SMA} and the IBM-1 with a term
quadratic in $(Q\otimes Q\otimes Q)_{0}$ \cite{6Q} (IBM-1) are also
shown.} \label{SJOs192}
\end{figure}

\begin{figure}[h]\centering
\includegraphics[width=80mm]{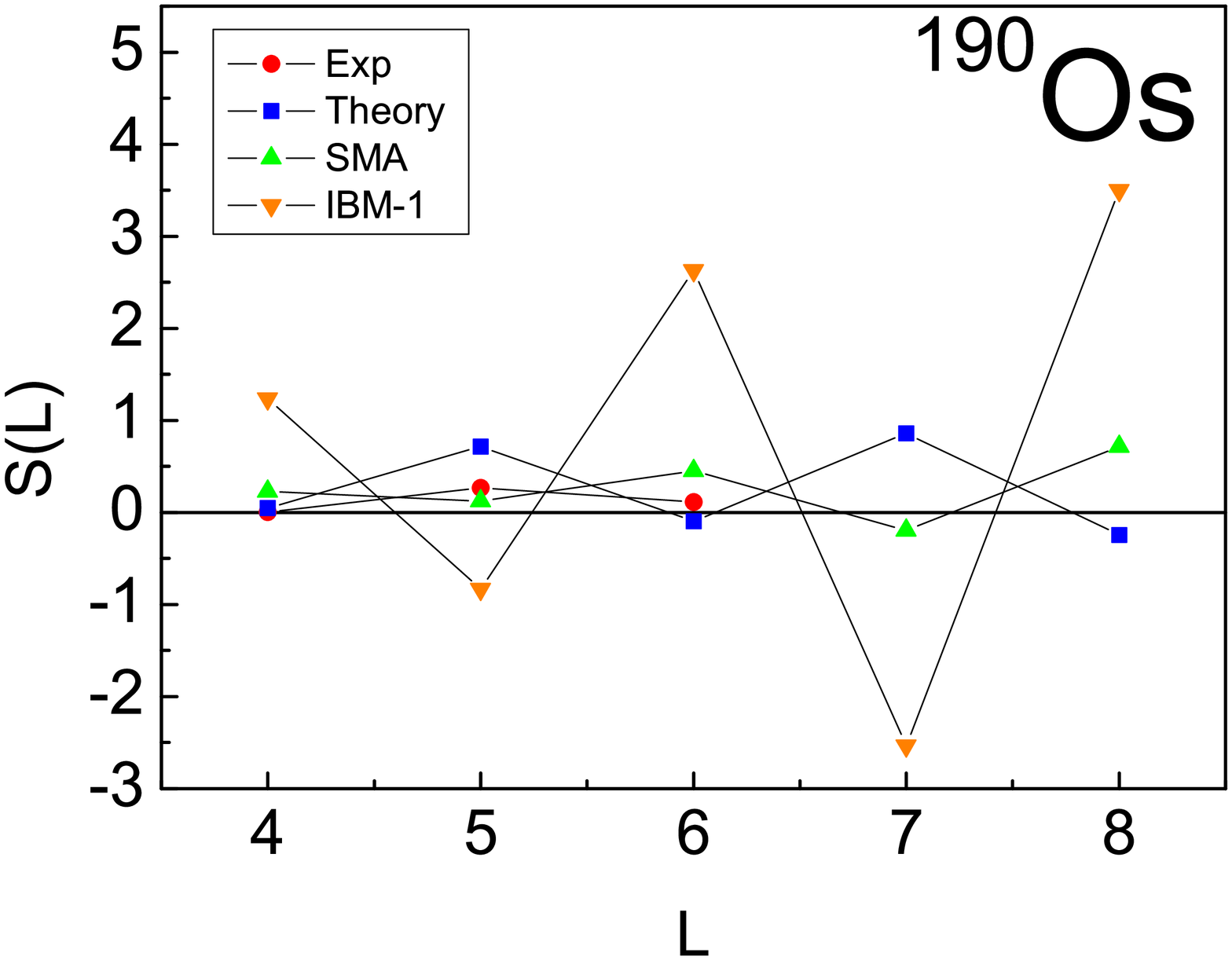}
\caption{(Color online) Calculated and experimental staggering of
the $\gamma$ band in $^{190}Os$. The predictions of the sextic and
Mathieu approach (SMA) \cite{SMA} and the IBM-1 with a term
quadratic in $(Q\otimes Q\otimes Q)_{0}$ \cite{6Q} (IBM-1) are also
shown.} \label{SJOs190}
\end{figure}

\begin{figure}[h]\centering
\includegraphics[width=80mm]{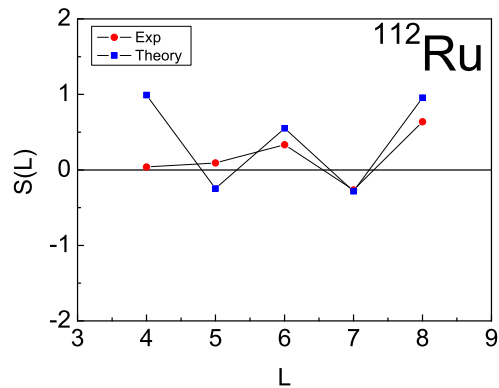}
\caption{(Color online) Calculated and experimental staggering of
the $\gamma$ band in $^{112}Ru$.} \label{SJRu112}
\end{figure}

Analysis of the experimental staggering in different isotopic chains
reveals several different patterns \cite{SJ} that can be categorized
based on the standard limits discussed in the IBM. Just to mention
few cases, the $Xe$, $Ba$ and $Ce$ nuclei are well-known examples
\cite{VibGS1},\cite{VibGS2},\cite{VibGS3} of the transition between
vibrational and $\gamma$-unstable structures that show strong
staggering with negative $S(L)$ values at even-$L$ and positive
$S(L)$ values at odd-$L$ spins. The heavy rare-earth nuclei $(N >
82)$, known to display an axially symmetric behavior, show a similar
staggering pattern with a smaller overall magnitude than that
observed in the $Xe$, $Ba$ and $Ce$ isotopes. Nuclei that display
staggering patterns very different from those described above are
scarce and include, for example, $^{192}Os$ and $^{112}Ru$. These
nuclei develop a staggering pattern where $S(L)$ is positive for
even-$L$ and negative for odd-$L$ values, i.e. with the opposite
phasing than in the other two cases mentioned above.

As shown in Ref.\cite{SJ} the geometrical models and the IBM-based
models can describe the basic trends observed in the experimental
staggering. It is shown that the geometrical models that incorporate
rigid triaxiality are characterized by strong staggering with
positive values for even-$L$ and negative values for odd-$L$ spins.
The staggering is largest for the RTRM where it increases linearly
with $L$ and smallest for the models that use a harmonic-oscillator
$\beta^{2}$ potential. Similarly, the IBM shows a jump over to the
triaxial region along the transition from $U(5)$ to $SU(3)$,
characterized by the same staggering pattern as the one found in the
geometrical models but with a smaller overall magnitude.

To see whether this signature is captured by the present approach,
we plotted the function $S(L)$ within the framework of the IVBM for
the nuclei under consideration in Figures
\ref{SJOs192}$-$\ref{SJRu112}, compared with the experimental data.
For $^{192}Os$ and $^{190}Os$, the predictions of the IBM-1 with a
term quadratic in $(Q\otimes Q\otimes Q)_{0}$ \cite{6Q} and sextic
and Mathieu approach (SMA) \cite{SMA} that incorporate
$\gamma$-rigid structures are also shown. As can be seen from the
figures, the present approach predicts a staggering pattern for
$^{192}Os$ and $^{112}Ru$ similar to the one observed in the
geometrical models and the IBM that incorporate triaxiality, and
$\gamma$-unstable type for $^{190}Os$, respectively.  For all nuclei
under considerations, the phases of the observed staggering patterns
are also correctly reproduced (in contrast to the SMA and IBM-1 in
the case of $^{190}Os$). For the two nuclei $^{192}Os$ and
$^{112}Ru$, the $\gamma$-rigid staggering is well developed in the
region $L \geq 5 $ where also its magnitude increase with the spin.
The latter suggests that the triaxiality evolves together with the
collectivity.

\subsection{Energy surfaces}

The geometry associated with a given Hamiltonian can be obtained by
the coherent state method. The standard approach to obtain the
geometrical properties of the system is to express the collective
variables in the intrinsic (body-fixed) frame of reference. Then the
ground-state energy surface is obtained by calculating the
expectation value of the boson Hamiltonian (\ref{HpM3}) with respect
to the corresponding coherent states. In the case of IVBM, the
(scaled) energy surfaces $\varepsilon(\rho,\theta)$ depend on two
coherent state parameters $\rho$ and $\theta$, determining the
"shape" of the nucleus \cite{TSIVBM},\cite{PSIVBM}. The latter are
related to the standard collective model "shape" parameters $\beta$
and $\gamma$. For more details we refer the reader to the
Refs.\cite{TSIVBM},\cite{PSIVBM}.


\begin{figure}[h]\centering
\includegraphics[width=43mm]{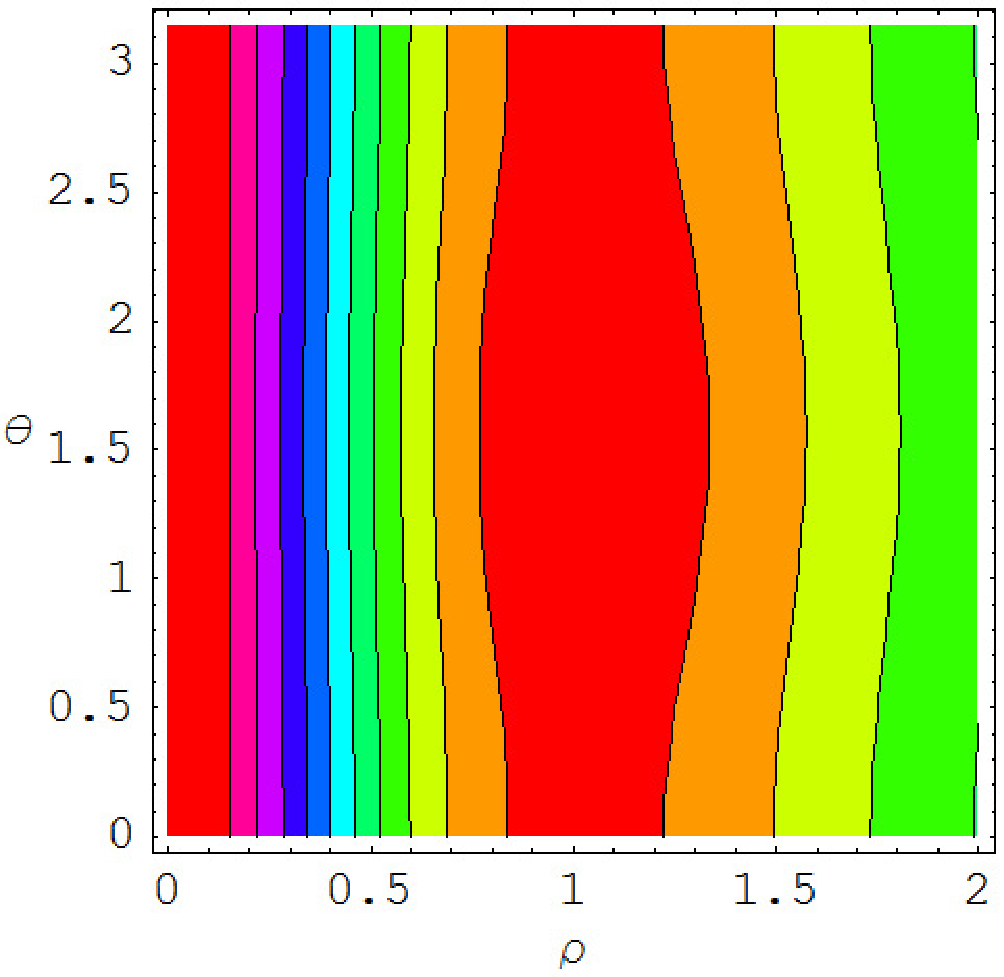}\hspace{10.mm}
\includegraphics[width=43mm]{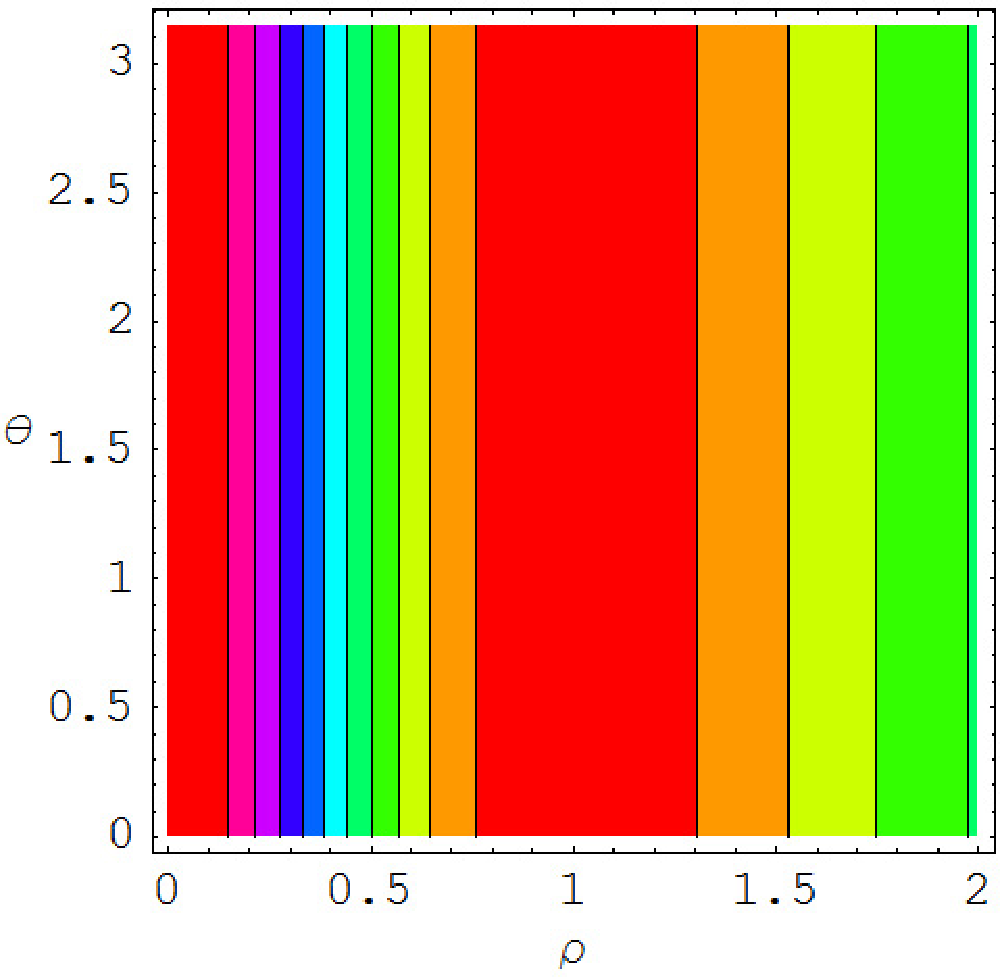}\hspace{10.mm}
\includegraphics[width=43mm]{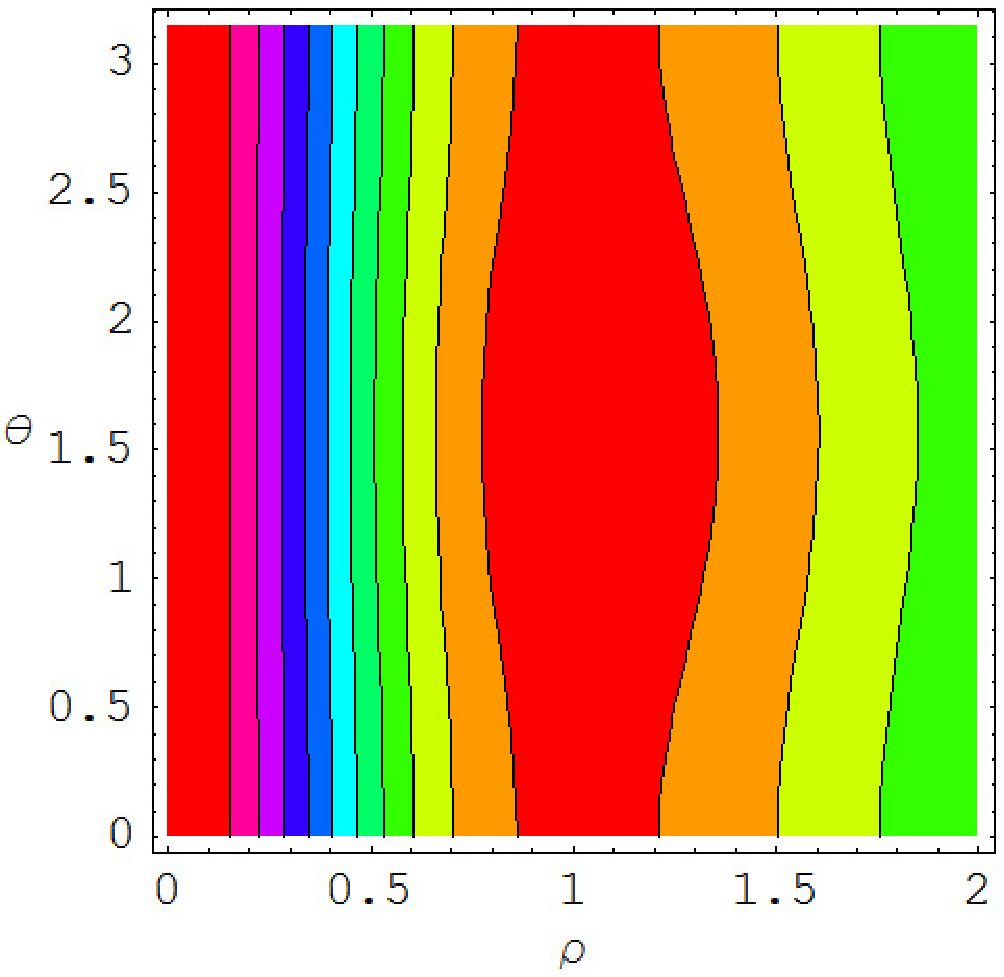}
\caption{(Color online) A contour plot of the scaled energy surfaces
$\varepsilon(\rho,\theta)$ corresponding to the Hamiltonian
(\ref{HpM3}) for $^{192}Os$, $^{190}Os$, and $^{112}Ru$,
respectively. Only the region $\rho > 0$ is depicted.} \label{2d}
\end{figure}


We plot the ground state energy surfaces in $^{192}Os$, $^{190}Os$,
and $^{112}Ru$ with the model parameters obtained in the fitting
procedure in the form of contour plots in Fig. \ref{2d}. From the
figure one sees nearly $\gamma$-flat potentials with very shallow
triaxial minima for the ground state in $^{192}Os$ and $^{112}Ru$,
while in $^{190}Os$ a typical for the $O(6)$ limit $\theta$-unstable
(or in IBM terms a $\gamma$-unstable) potential is observed, as
should be for nuclei that show strong staggering with negative
$S(L)$ values at even-$L$ and positive $S(L)$ values at odd-$L$
spins (see Fig.\ref{SJOs190}). The triaxial minima are obtained at
$\rho_{0} \approx 1$ and $\theta_{0}=90^{0}$ which corresponds to
$\gamma_{eff}=30^{0}$ \cite{TSIVBM}. In other words, the potentials
obtained in the present approach for $^{192}Os$ and $^{112}Ru$ are
indeed $\gamma$-dependent, representing the case of mixing of
$\gamma$-flat and $\gamma$-rigid structures.

Shallow triaxial minima in the energy surfaces of $^{190}Os$ and
$^{192}Os$ are predicted in recent HFB calculations with Gogny D1S
and Skyrme SLy4 forces \cite{Robledo}, but with an $O(6)$-like
doublet structure of the $\gamma$ bands. The IBM-2 energy surfaces
\cite{NomuraOs1}, \cite{NomuraOs2} derived from HFB ones with Gogny
D1S forces predict a $\gamma$-soft energy surface with a prolate
minimum for $^{190}Os$ and $\gamma$-unstable one for $^{192}Os$. The
HF and IBM-2 calculations with $SkM^{\ast}$ forces \cite{Nomura4}
produce a triaxial minimum and $\gamma$-soft energy surface with a
prolate minimum for $^{112}Ru$, respectively. A $\gamma$-soft energy
surface with a prolate minimum is predicted also in \cite{Aysto}. In
\cite{Troltenier}, the Generalized Collective Model including up to
sixth order (i.e. three-body) terms in $\alpha_{2\mu}$ able to
produce a stable triaxial minima predicts an energy surface with a
triaxial minimum for $^{112}Ru$.

\section{Transition probabilities}

It is known that the comparison of the experimental data with the
calculated transition probabilities is a more sensitive test for
each model under consideration. In this respect, in this section we
consider the electromagnetic properties of the IVBM with respect to
the reduction chain ($\ref{NDS}$).

The general approach required for the calculation of the matrix
elements of the transition operators defined in respect to the chain
($\ref{NDS}$), as well as its application to the more complicated
and complex problem of reproducing the $B(E2)$ transition
probabilities between the collective states of the ground band in
some isotopes exhibiting axially asymmetric features, is given in
\cite{tpu33}. Here, in addition to the intraband $E2$ transitions
between the states of the GSB we consider also the $E2$ transitions
between the states of the $\gamma$ band, as well as the interband
$M1$ transitions between the $\gamma$ and GSB in the two nuclei
$^{190}Os$ and $^{192}Os$, for which there is enough available
experimental data.

Transition probabilities are by definition $SO(3)$ reduced matrix
elements of transition operators $T^{E2}$ between the $|i\rangle
-$initial and $|f\rangle -$final collective states
\begin{equation}
B(E2;L_{i}\rightarrow L_{f})=\frac{1}{2L_{i}+1}\mid \langle \quad
f\parallel T^{E2}\parallel i\quad \rangle \mid ^{2}. \label{deftrpr}
\end{equation}
Using the tensorial properties of the $Sp(12,R)$ generators and the
mapping considered above, it is easy to define the $E2$ transition
operator between the states of the GSB and $\gamma$ band as
\cite{tpu33}:
\begin{align}
T^{E2} = \ &e[A'^{20}_{[210]_{3}} +\theta ([F\times F]^{\quad \quad
\quad \quad 20}_{[2]_{3}[2]^{\ast}_{3}\quad [420]_{3}}\notag\\
\notag\\
&+[G\times G]^{\quad \quad \quad \quad
20}_{[2]^{\ast}_{3}[2]_{3}\quad [420]_{3}})], \label{te2}
\end{align}
where the first tensor operator is the $SU^{\ast}(3)$ quadrupole
operator and actually changes only the angular momentum with $\Delta
L=2$ within a given irrep $(\lambda,\mu)$.

The tensor product
\begin{align}
&[F\times F]^{\quad \quad \quad \quad
20}_{[2]_{3}[2]^{\ast}_{3}\quad [420]_{3}} \notag \\
\notag \\
&=\sum C^{[420]_{3}}_{[2]_{3},[2]^{\ast}_{3}} C^{(2,0) (0,2)
(2,2)}_{\quad 2 \quad 2 \quad 2} C^{20}_{20,20} \notag \\
\notag \\
&\times F^{\quad \ \quad \quad 20}_{[2]_{3}[0]^{\ast}_{3}\quad
[2]_{3}} F^{\quad \ \quad \quad 20}_{[0]_{3}[2]^{\ast}_{3}\quad
[-2]_{3}} \label{FF}
\end{align}
of the rasing generators of $Sp(12,R)$ changes the number of bosons
by $\Delta N=4$ and $\Delta L=2$.

\begin{figure}[h]\centering
\includegraphics[width=80mm]{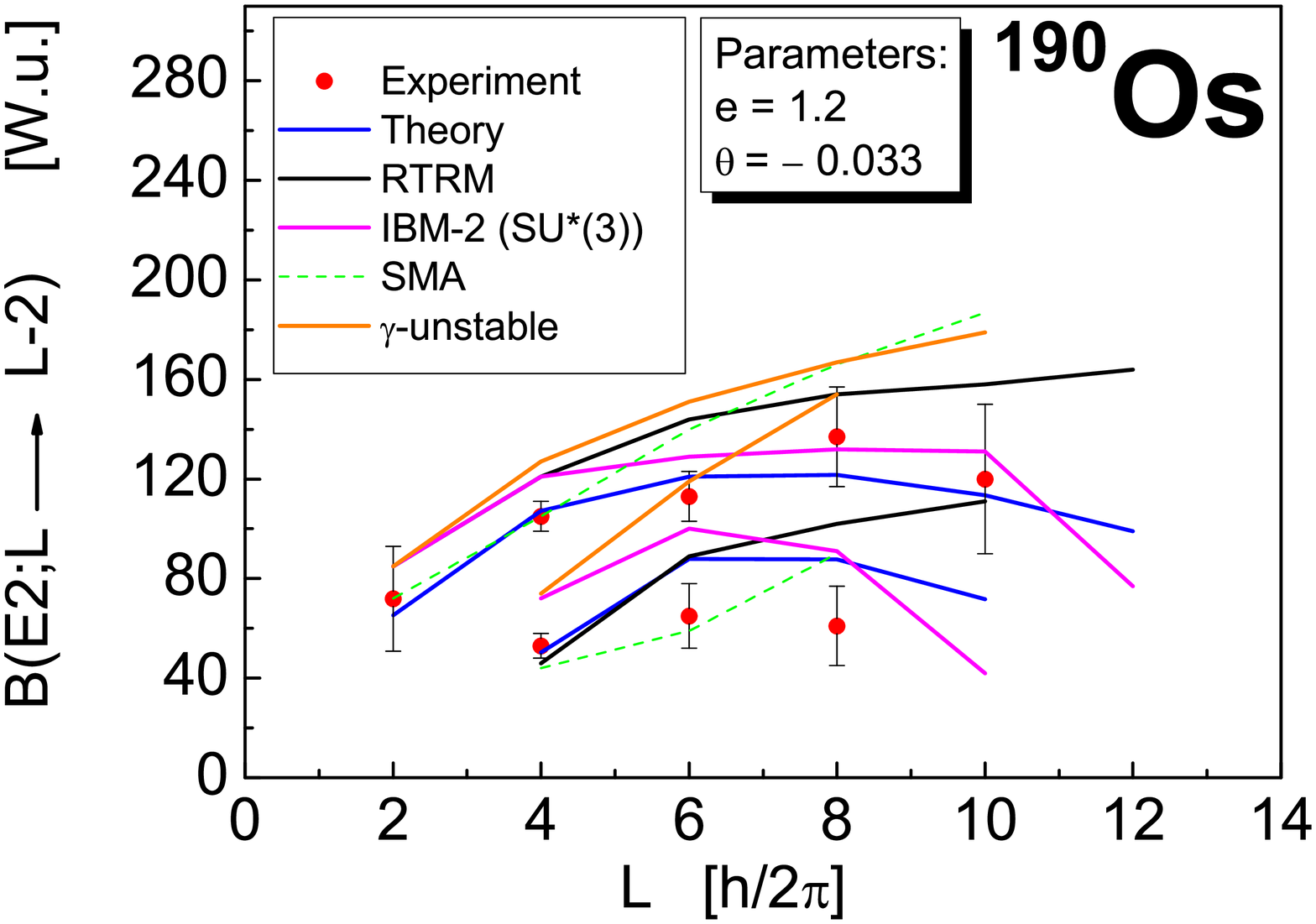}\hspace{1.mm}
\includegraphics[width=80mm]{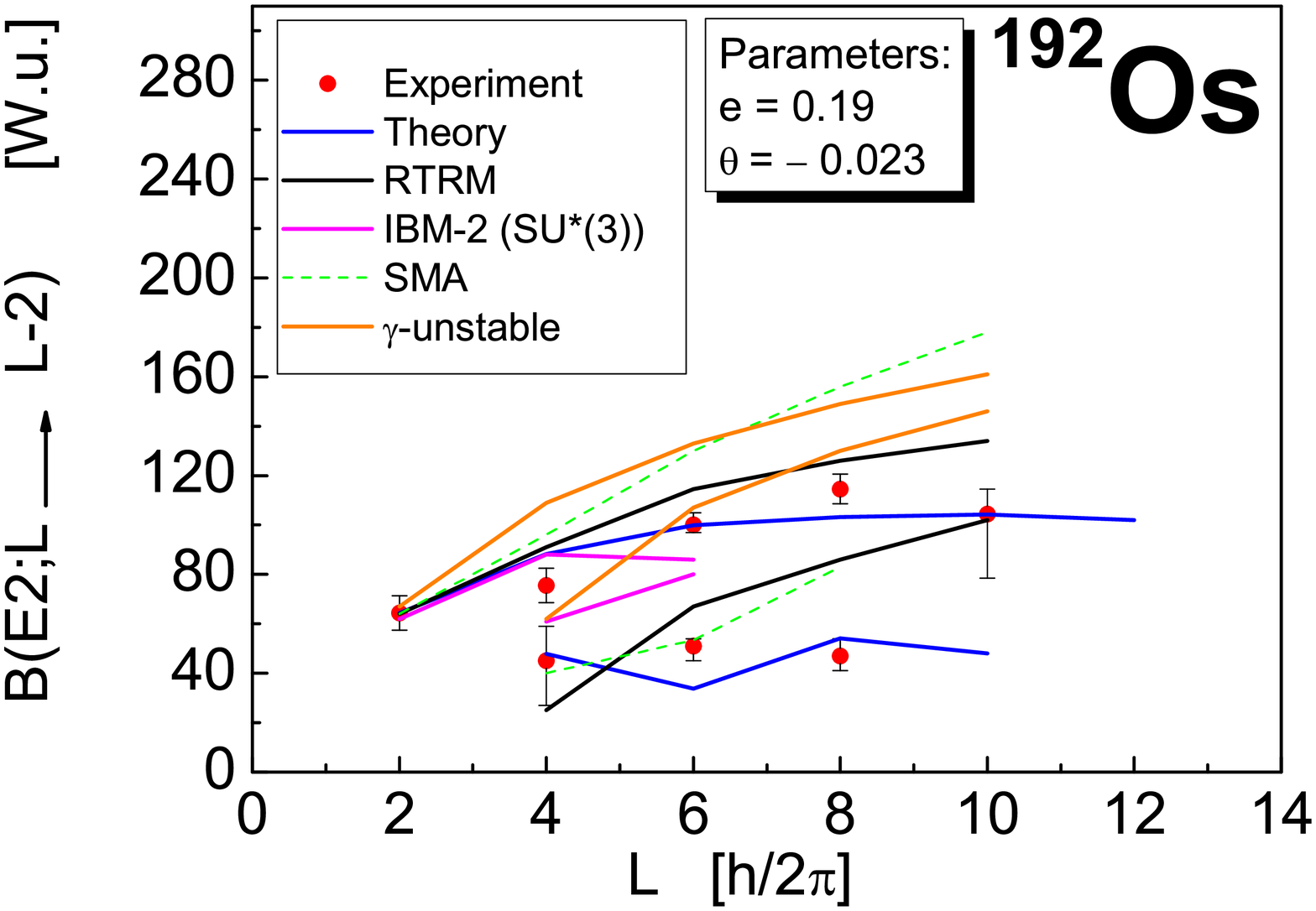}
\caption{(Color online) Comparison of theoretical and experimental
values for the $B(E2)$ transition probabilities between the
collective states of the ground state band and $\gamma$ band in
$^{190}Os$ and $^{192}Os$, respectively. The theoretical results of
some other collective models which accommodate the $\gamma$-rigid or
$\gamma$-unstable structures are also shown.} \label{BE2}
\end{figure}

The application actually consists of fitting the two parameters $e$
and $\theta$ of the transition operator $T^{E2}$ (\ref{te2}) to
experiment for each isotope. The experimental data \cite{exp} for
the $B(E2)$ transition probabilities between the states of the GSB
and $\gamma$ band are compared with the corresponding theoretical
results of the symplectic IVBM in Figure \ref{BE2}. For comparison,
the theoretical predictions of IBM-2 \cite{Walet} in its
$SU^{\ast}(3)$ limit, RTRM \cite{Toki}, sextic and Mathieu approach
(SMA) \cite{SMA}, and $\gamma$-unstable model of Wilets and Jean
\cite{WJ} are also shown. From the figure one can see that all
models presented reproduce the general trend of the experimental
data, but nevertheless with the increasing of spin the RTRM, SMA,
and $\gamma$-unstable model overestimate the observed experimental
data well described by the IVBM. Note different values obtained for
the model parameters for the two cases -with ($^{192}Os$) and
without ($^{190}Os$) mixing of the $SU(3)$ multiplets.

\begin{figure}[h]\centering
\includegraphics[width=80mm]{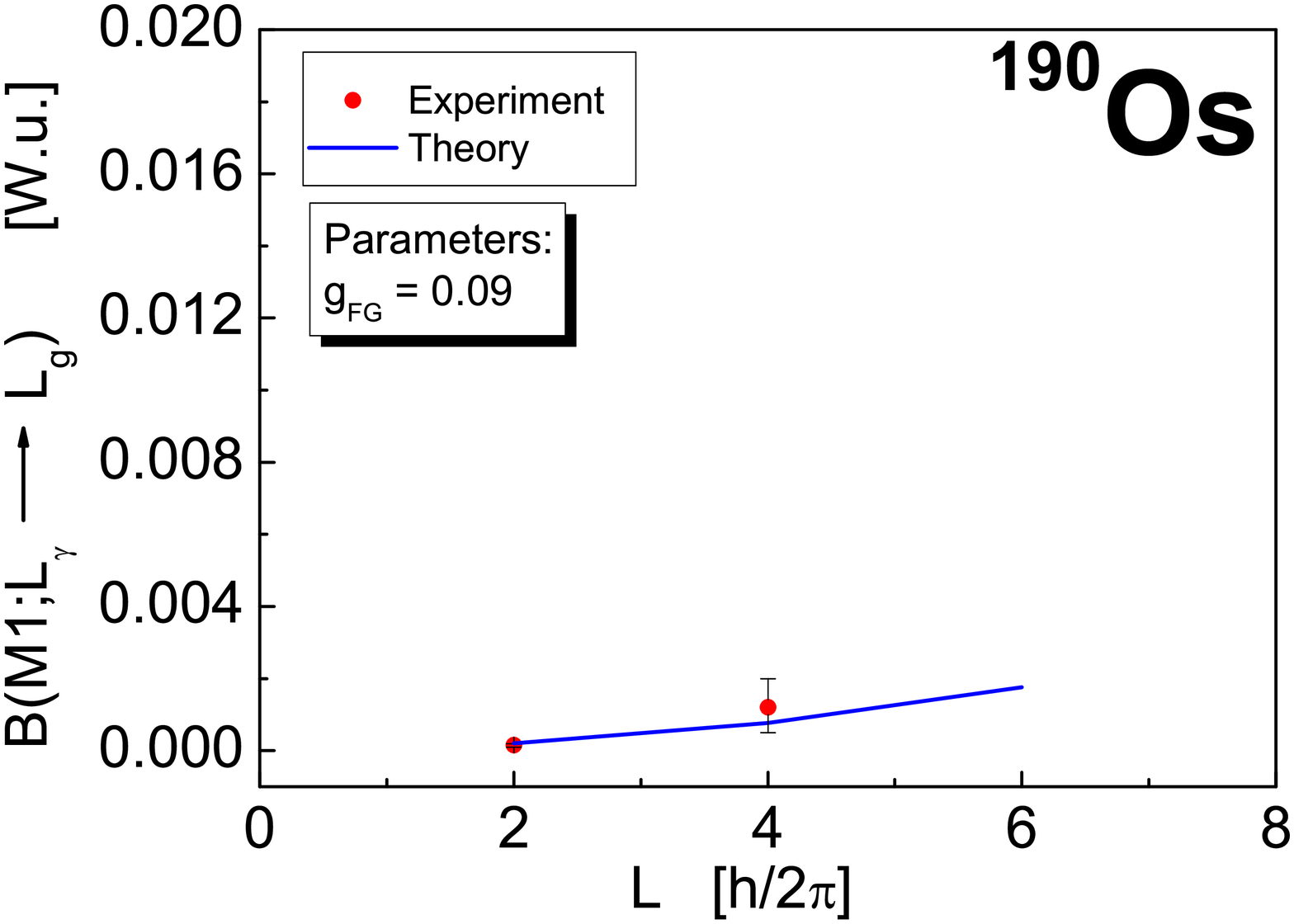}\hspace{1.mm}
\includegraphics[width=80mm]{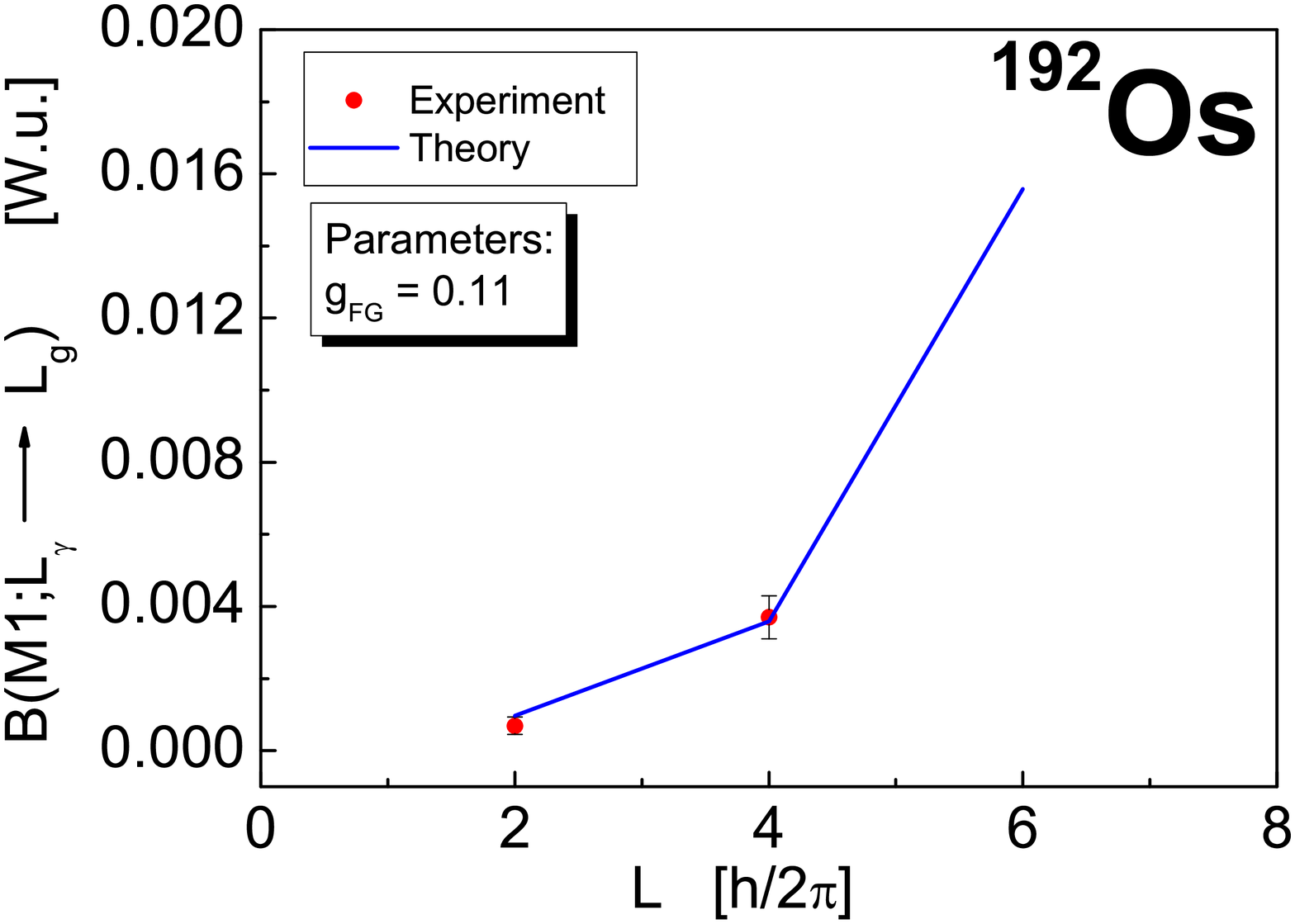}
\caption{(Color online) Comparison of theoretical and experimental
values for the interband $B(M1)$ transition probabilities between
the ground state band and $\gamma$ band in $^{190}Os$ and
$^{192}Os$, respectively.} \label{BM1}
\end{figure}

Next, we consider the interband $M1$ transitions between the GSB and
$\gamma$ band for which there is a very few experimental data. Using
the tensorial properties of the basis states and the indetification
of the experimentally observed states of the GSB and $\gamma$ band
with the subset of stretched states from the $\nu = 0$ and $\nu =
-2$ irreps of $U(3,3)$, respectively, we can define as an
appropriate the following $M1$ transition operator lying in the
enveloping algebra of $Sp(12,R)$:

\begin{equation}
T^{M1} = g_{FG}([L\times F]^{\ 10}_{[4,3,0]_{3}}+[L\times G]^{\
10}_{[4,1,0]_{3}}), \label{tm1}
\end{equation}
where the operators $F$, $G$ and $L$ are defined by Eqs.
($\ref{Fs}$), ($\ref{Gs}$) and($\ref{LM}$), respectively. (For more
details concerning the tensor properties of the $Sp(12,R)$ algebra
generators and the construction of the symplectic basis see
Ref.\cite{tpu33}.)

In Figure \ref{BM1} we compare our theoretical predictions for the
interband M1 transitions bitween the states of the GSB and $\gamma$
band with experiment \cite{exp} for the two nuclei $^{190}Os$ and
$^{192}Os$, respectively. Despite the very poor experimental
information, one sees a good reproduction of the $B(M1)$ transition
probabilities for the two nuclei under consideration.

The results obtained for the transition probabilities considered in
this section also prove the correct mapping of the basis states to
the experimentally observed ones and reveal the relevance of the
IVBM in the description of nuclei that exhibit axially asymmetric
features in their spectra. Certainly, the used dynamical symmetry of
the IVBM needs to be further explored by more detailed comparison of
its predictions with the available experimental data.

\section{Summary}

In the present work, we apply one of the dynamical symmetry limits
of the two-fluid Interacting Vector Boson Model, defined through the
chain $Sp(12,R) \supset U(3,3) \supset U_{p}(3) \otimes
\overline{U_{n}(3)} \supset SU^{\ast}(3) \supset SO(3)$, for the
description of some even-even nuclei, possessing axial asymmetry. We
have investigated the effect of the introduction of a Majorana
interaction to the $SU^{\ast}(3)$ model Hamiltonian on the
excitation energies of the ground and $\gamma$ bands. The latter
introduces a potential which has a minimum at $\gamma = 30^{0}$ and
change the $\gamma$-band doublet structure from that of
$\gamma$-unstable to that of $\gamma$-rigid type. This allows for
the description of these two limiting cases, as well as the
situation in between, which is characterized by more uniform energy
level spacings in the $\gamma$-band, and described actually by
$\gamma$-dependent potentials.

The theoretical predictions are compared with the experimental data
for $^{192}Os$, $^{190}Os$, and $^{112}Ru$ isotopes. It is shown
that by taking into account the full symplectic structures in the
considered dynamical symmetry of the IVBM, the proper description of
the energy spectra and the $\gamma$-band energy staggering of the
nuclei under considerations can be achieved. The obtained results
show that the potential energy surfaces for the following two nuclei
$^{192}Os$ and $^{112}Ru$, possess almost $\gamma$-flat potentials
with very shallow triaxial minima, suggesting a more complex and
intermediate situation between $\gamma$-rigid and $\gamma$-unstable
structures.

Symplectic dynamical symmetries provide for larger classification
spaces and richer subalgebraic structures allowing to incorporate
more complicated structural effects in nuclear spectra.
Additionally, the algebraic structure of the model allows both the
basis states and the transition operators to be defined as tensors
in respect to the considered dynamical symmetry which simplifies the
calculation of the transition probabilities by using a generalized
version of the Wigner-Eckart theorem. This allows the model to be
further tested on the more complicated problem of reproducing the
transition probabilities between the collective states attributed to
the basis states of the Hamiltonian. Exploiting this, the theory is
further applied for the calculation of the absolute $B(E2)$
intraband transition probabilities between the states of the ground
state band and $\gamma$ band, as well as the $B(M1)$ interband
transition probabilities between the states of the GSB and $\gamma$
band for the two nuclei $^{192}Os$ and $^{190}Os$ for which there is
enough available experimental data. The theoretical predictions are
compared with experiment and for the $B(E2)$ values with the
predictions of some other collective models incorporating the
$\gamma$-rigid or $\gamma$-unstable structures.

The results obtained for the energy levels, the energy staggering of
the $\gamma$ band and the transition probabilities of the considered
nuclei prove the correct mapping of the basis states to the
experimentally observed ones and reveal the relevance of the used
dynamical symmetry of IVBM in the description of nuclei exhibiting
axially asymmetric features in their spectra.




\begin{thebibliography}{}

\bibitem{ZC1} N. V. Zamfir and R. F. Casten, Phys. Lett. \textbf{B260}, 265
(1991).

\bibitem{SJ} E. A. McCutchan, D. Bonatsos, N. V. Zamfir, and R. F.
Casten, Phys. Rev. \textbf{C76}, 024306 (2007).

\bibitem{WJ} L. Wilets and M. Jean, Phys. Rev. \textbf{102}, 788 (1956).

\bibitem{DF} A. S. Davydov and G. F. Filippov, Nucl. Phys. \textbf{8}, 237
(1958).

\bibitem{SSa} Liao Ji-zhi, Phys. Rev. \textbf{C51}, 141 (1995).

\bibitem{SSb}  C.Bihari \emph{et al.},
Phys. Scr. \textbf{77}, 055201 (2008).

\bibitem{SSc} C. Bihari \emph{et al.},
Phys. Scr. \textbf{78}, 045201 (2008).

\bibitem{SSd} Mani Varshney,
Phys. Scr. \textbf{83}, 015201 (2011).

\bibitem{TSIVBM} H. G. Ganev, Phys. Rev. \textbf{C84}, 054318 (2011).

\bibitem{BargMosh1} V. Bargmann and M. Moshynsky,
Nucl. Phys. \textbf{18}, 697 (1960).

\bibitem{BargMosh2} V. Bargmann and M. Moshynsky,
Nucl. Phys. \textbf{23}, 177 (1961).

\bibitem{PSIVBM} H. G. Ganev, Phys. Rev. \textbf{C83}, 034307 (2011).

\bibitem{GGG} H. Ganev, V. P. Garistov, and A. I. Georgieva,
Phys. Rev. C \textbf{69}, 014305 (2004).

\bibitem{str} D. J. Rowe, Rep. Prog. Phys. \textbf{48}, 1419
(1985).


\bibitem{exp} National Nuclear Data Center (NNDC),
http://www.nndc.bnl.gov/



\bibitem{NomuraOs1} K. Nomura \emph{et al.},
Phys. Rev. \textbf{C 83}, 054303 (2011).

\bibitem{NomuraOs2} K. Nomura \emph{et al.},
Phys. Rev. \textbf{C 84}, 054316 (2011).

\bibitem{OsTri1} N. Redon \emph{et al.}, Phys. Lett. \textbf{B181}, 223
(1986).

\bibitem{OsTri2} W. Boeglin \emph{et al.},
Nucl. Phys. \textbf{A 477}, 289 (1988).

\bibitem{OsTri3} P. Sarriguren, R. Rodr´ýguez-Guzm´an, L. M.
Robledo, Phys. Rev. \textbf{C 77}, 064322 (2008).

\bibitem{OsIBM} R. F. Casten and J. A. Cizewski, Nucl. Phys. \textbf{A 309}, 477 (1978).

\bibitem{Casten} R. F. Casten, \emph{Nuclear Structure from a Simple Perspective} (Oxford
University, Oxford, 1990).

\bibitem{VibGS1} R.F. Casten, P. Von Brentano, K. Heyde, P. Van Isacker,
J. Jolie, Nucl. Phys. \textbf{A 439}, 289 (1985).

\bibitem{Bonche} P. Bonche \emph{et al.},
Nucl. Phys. \textbf{A500}, 308 (1989).

\bibitem{CCQH} L. Fortunato, C. E. Alonso, J. M. Arias, J. E.
Garcia-Ramos, and A. Vitturi, Phys. Rev. \textbf{C 84}, 014326
(2011).

\bibitem{Robledo} L. M. Robledo, R. Rodriguez-Guzman, and P.
Sarriguren, J. Phys. G: Nucl. Part. Phys. \textbf{36}, 115104
(2009).

\bibitem{Nomura2} K. Nomura \emph{et al.},
Phys. Rev. \textbf{C 83}, 014309 (2011).

\bibitem{Nomura3} K. Nomura \emph{et al.},
Phys. Rev. \textbf{C 84}, 014302 (2011).


\bibitem{IBMRu} A. Frank, P. Van Isacker, and P. D. Warner,
Phys. Lett. \textbf{B 197}, 474 (1987).

\bibitem{GCMRu} D. Troltenier et al.,
Z. Phys.  \textbf{A 338}, 261 (1991).

\bibitem{RTRMRu1} J. Stachel et al, Z. Phys.
\textbf{A316}, 105 (1984).

\bibitem{RTRMRu2} J. A. Shannon et al., Phys. Lett.
\textbf{B 336}, 136 (1994).

\bibitem{VibGS2} G. Puddu, O. Scholten, T. Otsuka, Nucl. Phys. \textbf{A 348}, 109
(1980).

\bibitem{VibGS3} N.V. Zamfir, R.F. Casten, Phys. Lett. \textbf{B 152}, 22 (1985).

\bibitem{6Q} G. Thiamova, Eur. Phys. J. \textbf{A 45}, 81 (2010).

\bibitem{SMA} A. A. Raduta and P. Buganu, Phys. Rev.
\textbf{C 83}, 034313 (2011).

\bibitem{Nomura4} K. Nomura \emph{et al.},
Phys. Rev. \textbf{C 81}, 044307 (2010).


\bibitem{Aysto} J. Aysto \emph{et al.},
Nucl. Phys. \textbf{A515}, 365 (1990).

\bibitem{Troltenier} D. Troltenier \emph{et al.},
Nucl. Phys. \textbf{A601}, 56 (1996).

\bibitem{tpu33} H. G. Ganev, Phys. Rev. \textbf{C 86}, 054311 (2012).



\bibitem{Walet} N. R. Walet and P. J. Brussaard,
Nucl. Phys. \textbf{A474}, 61 (1987).

\bibitem{Toki} H. Toki and A. Faessler,
Z. Phys.  \textbf{A 276}, 35 (1976).



\end{thebibliography}
\end{document}